\documentclass{PoS}

\usepackage{amsmath}
\usepackage{paralist}
\usepackage{sidecap}
\usepackage[utf8]{inputenc}

\title{Systematics in nucleon matrix element calculations}

\ShortTitle{Systematics in nucleon matrix element calculations}

\author{\speaker{Jeremy Green}\\
        NIC, Deutsches Elektronen-Synchrotron, 15738 Zeuthen, Germany\\
        E-mail: \email{jeremy.green@desy.de}}

      \abstract{The current status of calculations of simple nucleon
        structure observables is reviewed, with a focus on the axial
        charge. A major challenge is the combination of an
        exponentially decaying signal-to-noise ratio and the need for
        large source-sink separations to eliminate excited-state
        contributions; efforts to understand and deal with this
        problem are the focus of the largest section of this
        review. Finite-volume effects and chiral extrapolation are
        also briefly discussed.}

\FullConference{The 36th Annual International Symposium on Lattice Field Theory - LATTICE2018\\
		22-28 July, 2018\\
		Michigan State University, East Lansing, Michigan, USA.}

\begin{document}

\section{Introduction}

The structure of nucleons has been studied extensively in experiments,
and nucleons also play a vital role as experimental probes. This makes
the controlled study of nucleons using lattice QCD a very attractive
goal. Some of the simplest matrix elements of interest include the
scalar and tensor charges, which control BSM contributions to neutron
beta decay, and the sigma terms, which control the sensitivity of dark
matter detection to WIMPs.

In order to make a significant impact, lattice calculations of nucleon
structure must become precision calculations, with full control over
all systematics. The traditional test observable is the isovector
axial charge $g_A$:
\begin{equation}
  \left\langle p(P,s'\middle)|\bar u\gamma^\mu \gamma_5 d\middle|n(P,s)\right\rangle
\equiv g_A \bar u_p(P,s') \gamma^\mu \gamma_5 u_n(P,s).
\end{equation}
It is simple to compute, being an isovector forward matrix element,
and it is measured precisely in beta decay experiments\footnote{It
  should be noted, however, that over time the experimental value has
  drifted upward slightly; see the PDG's history plot.}, with the
latest PDG value~\cite{Tanabashi:2018oca} being
$g_A=1.2724(23)$. However, calculating it accurately has been
challenging: there is a long history of results being below the
experimental value. Because systematic effects in $g_A$ are
significant and it has been studied the most extensively, the axial
charge will be the exclusive focus of this review. This review will
also focus on recent results, particularly those presented at this
conference.

This review is organized as follows. Section~\ref{sec:methods} is a
brief overview of how matrix elements are computed. Excited-state
effects, which are a particularly challenging source of systematic
uncertainty, are discussed at length in
Section~\ref{sec:exc}. Finite-volume effects and dependence on the
pion mass are briefly reviewed in Sections~\ref{sec:volume} and
\ref{sec:chiral}. Finally, a summary and outlook is given in
Section~\ref{sec:outlook}.

\section{Methodology of matrix elements}\label{sec:methods}

The simplest approach for computing the forward hadronic matrix
element of operator $\mathcal{O}$ is to use a single interpolator
$\chi$ at zero momentum. One computes two-point and three-point
functions and performs a spectral decomposition,
\begin{gather}
 \begin{aligned}
   C_\text{2pt}(t) &\equiv \left\langle \chi(t) \chi^\dagger(0) \right\rangle & \qquad
   C_\text{3pt}(\tau,T) &\equiv \left\langle \chi(T) \mathcal{O}(\tau) \chi^\dagger(0) \right\rangle \\
   &= \sum_n |Z_n|^2 e^{-E_n t}, &
   &= \sum_{n,n'} Z_{n'} Z_n^* \langle n'|\mathcal{O}|n\rangle
     e^{-E_n\tau} e^{-E_{n'}(T-\tau)},
 \end{aligned}
\end{gather}
where $Z_n\equiv\langle\Omega|\chi|n\rangle$ is the overlap of the
interpolator onto state $n$. In the limit where all time separations
$t$, $\tau$, and $T-\tau$ are large, the ground state dominates:
\begin{align}
  C_\text{2pt}(t) &\to |Z_0|^2 e^{-E_0 t}\left(1 + O(e^{-\Delta E t})\right),\\
  C_\text{3pt}(\tau,T) &\to |Z_0|^2 e^{-E_0 t}\left( \langle 0|\mathcal{O}|0\rangle
  + O(e^{-\Delta E \tau}) + O(e^{-\Delta E(T-\tau)} + O(e^{-\Delta E T}) \right),
\end{align}
where $\Delta E\equiv E_1-E_0$ is the energy gap to the first excited
state.

In the \textbf{ratio method}, the prefactors are cancelled to obtain
the ground-state matrix element:
\begin{equation}
  R(\tau,T) \equiv \frac{C_\text{3pt}(\tau,T)}{C_\text{2pt}(T)}
  \to \langle 0|\mathcal{O}|0\rangle + O(e^{-\Delta E\tau}) + O(e^{-\Delta E(T-\tau)})
  + O(e^{-\Delta E T}).
\end{equation}
For forward matrix elements, the excited-state contributions are
symmetric about $\tau=T/2$. They can be minimized by placing the
operator at the midpoint, yielding $R(\tfrac{T}{2},T) = \langle
0|\mathcal{O}|0\rangle + O(e^{-\Delta E T/2})$.

\begin{figure}
  \centering
  \includegraphics[width=0.7\textwidth]{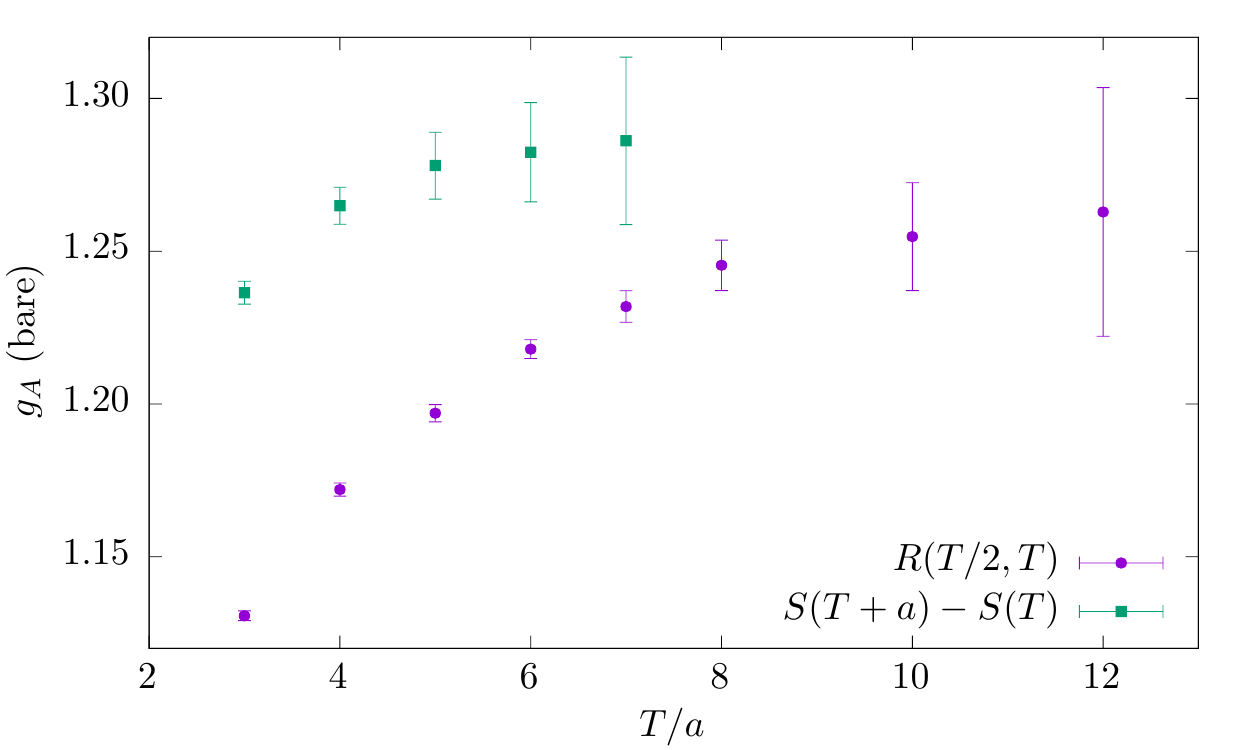}
  \caption{Bare axial charge, determined using the ratio method
    (purple circles) and summation method (green squares), versus
    source-sink separation. This calculation was performed at the
    physical pion mass with lattice spacing
    $a=0.116$~fm~\cite{Hasan_inprep}. Note that statistics for
    $T/a\geq 6$ are doubled compared with $T/a<6$.}
  \label{fig:ratio_summ}
\end{figure}

An alternative approach is the \textbf{summation
  method}~\cite{Maiani:1987by,Gusken:1989ad}, which involves summing
over the operator insertion time $\tau$. The terms that are
independent of $\tau$ grow linearly with the source-sink separation
$T$ in the sum, whereas the terms that have an exponential dependence
on $\tau$ produce partial sums of geometric series. The derivative of
the sum yields the ground-state matrix element:
\begin{equation}
  S(T) \equiv a\sum_\tau R(\tau,T),\quad
  \frac{d}{dT}S(T) = \langle 0|\mathcal{O}|0\rangle + O(T e^{-\Delta E T}).
\end{equation}
There is some flexibility about the interval over which $\tau$ is
summed. One option is to use the ``interior'' region where
$\mathcal{O}$ is between the source and the sink, possibly excluding a
fixed number of points at each end so that the sum is from $\tau_0$ to
$T-\tau_0$. Another option is to sum over the whole lattice, which is
the method that was first used. In two talks at the Lattice 2010
conference~\cite{Capitani:2010sg,Bulava:2010ej}, it was pointed out
that contributions from excited states decay more rapidly than for the
ratio-midpoint method; this revived interest in the summation
method. In practice, the summation method has been found to produce a
larger statistical uncertainty than the ratio method, which can negate
some of its advantage; see Fig.~\ref{fig:ratio_summ}.

In the last few years, there has been some use of methodologies based
on the Feynman-Hellmann theorem. This states that a matrix element in
a given state can be obtained from the derivative of the state's
energy with respect to a perturbation in the Lagrangian:
\begin{equation}\label{eq:FH}
  \mathcal{L}(\lambda)\equiv \mathcal{L} + \lambda\mathcal{O}
\implies
\left.\frac{\partial}{\partial\lambda}E_n(\lambda)\right|_{\lambda=0}
= \langle n|\mathcal{O}|n\rangle.
\end{equation}
Discrete derivatives are sometimes used, particularly for the nucleon
sigma terms, where the theorem relates nucleon scalar matrix elements
to derivatives of the nucleon mass with respect to quark
masses. Evaluating the derivative of a two-point function exactly
leads directly to the summation method:
\begin{equation}
  -\left.\frac{\partial}{\partial\lambda}\log C_\text{2pt}(t)\right|_{\lambda=0}
 = S(t),
\end{equation}
where the sum is taken over all timeslices. In the large-$t$ limit,
the time derivative of this equation yields Eq.~\eqref{eq:FH} for the
ground state. This result appeared in the original summation-method
paper~\cite{Maiani:1987by} and has been rederived in recent
years~\cite{Chambers:2014qaa,Savage:2016kon,Bouchard:2016heu}.

\section{Excited-state contamination}\label{sec:exc}

Unwanted contributions from excited states decay exponentially and
will be highly suppressed if $\Delta E T \gg 1$. However, the
signal-to-noise problem~\cite{Lepage:1989hd} prevents calculations
from being performed at large $T$: this ratio decays as
$e^{-(E_0-\tfrac{3}{2}m_\pi)T}$. To be more concrete about the ``brute
force'' approach of simply using a large source-sink separation,
assume that we are using the ratio method in the asymptotic regime,
where the statistical errors and excited-state contributions scale as
\begin{equation}
  \delta_\text{stat}\propto N^{-1/2}e^{(m_N-\tfrac{3}{2}m_\pi)T},\quad
  \delta_\text{exc}\propto e^{-\Delta E T/2},
\end{equation}
respectively, where $N$ is the number of statistical
samples. Supposing that we want these two uncertainties to scale
together, i.e.\ $\delta_\text{exc}=\alpha\delta_\text{stat}\equiv
\delta$ for some constant $\alpha$, then as $\delta$ is decreased the
source-sink separation must be increased to suppress excited state
effects. An increase in statistics is also required, both to
compensate for the reduced signal-to-noise ratio, and also to meet the
smaller target statistical error; the required statistics are given by
\begin{equation}
  N \propto \delta^{-\left(2+\frac{4m_N-6m_\pi}{\Delta E}\right)}.
\end{equation}
At the physical pion mass with $\Delta E=2m_\pi$ (see the next
subsection), the exponent is roughly $-13$, much larger than the $-2$
that is obtained when neglecting excited states. This situation could
be significantly improved if multilevel
methods~\cite{Ce:2016idq,Ce:2016ajy} or other
ideas~\cite{Detmold:2018eqd} are able to reduce the signal-to-noise
problem.

Because of the difficulty in going to large $T$, there has been much
effort spent on removing excited states from data at relatively small
source-sink separations: the two main approaches are improving the
interpolating operator and modeling the excited-state contributions.

\subsection{Theoretical expectations}

\begin{figure}
  \centering
  \includegraphics[width=0.495\textwidth]{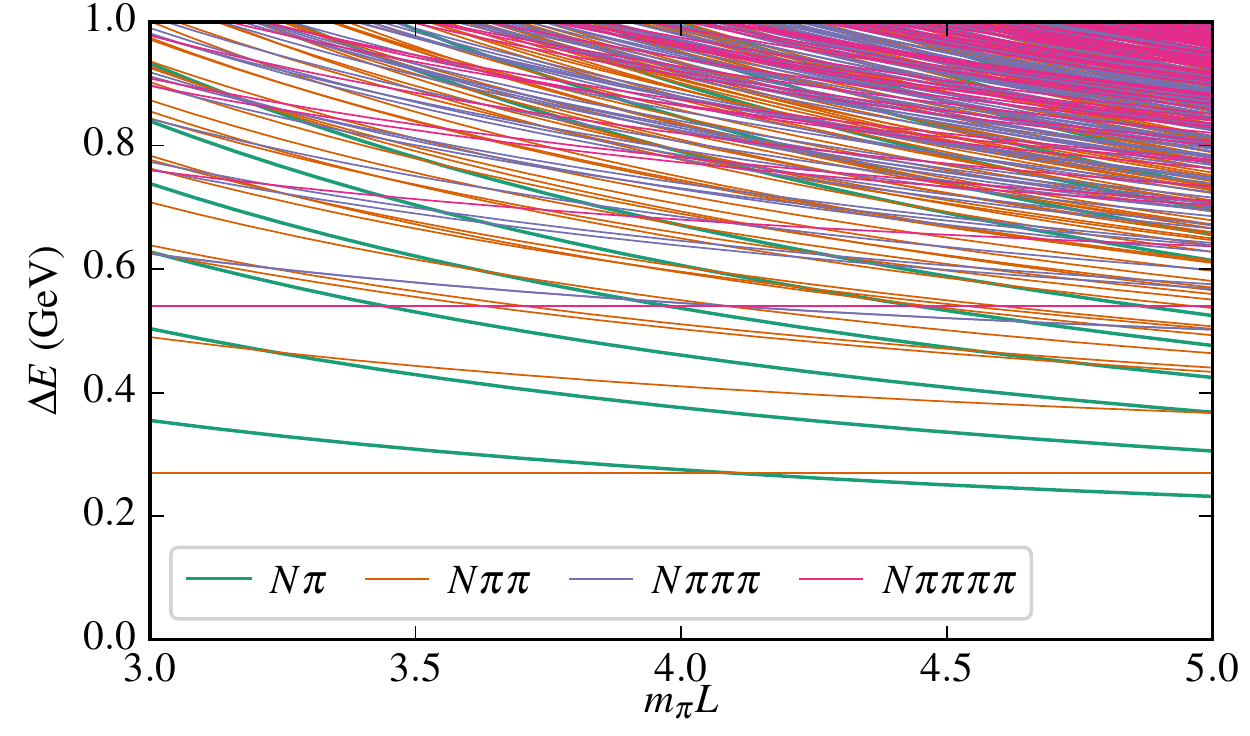}
  \includegraphics[width=0.495\textwidth]{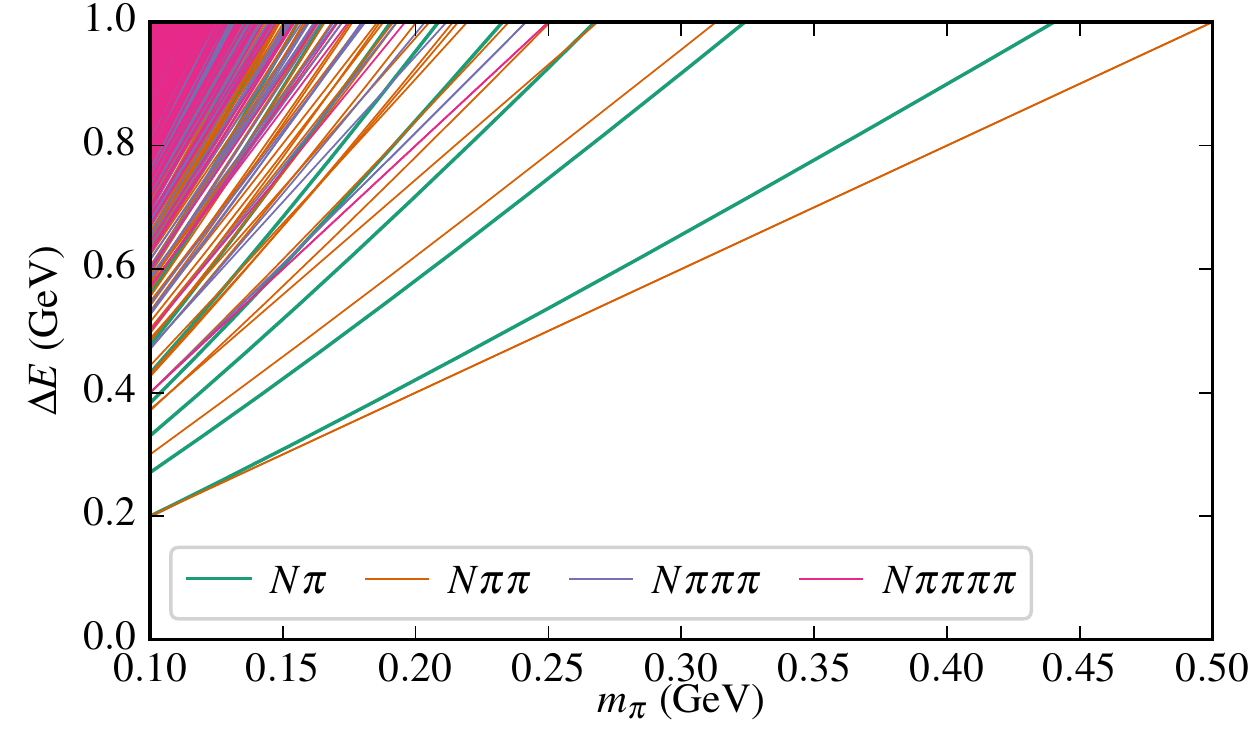}
  \caption{Energy gaps between noninteracting multiparticle excited
    states states and the nucleon ground state. Left: as a function of
    box size, at the physical pion mass. Right: versus pion mass, for
    $m_\pi L=4$. At heavier pion masses one also expects additional
    relatively low-lying states as resonances become stable; some
    possibilities include $N^*(1440)$, $\Delta\pi$, and $N\sigma$.}
  \label{fig:dE_free}
\end{figure}

An approximation to the finite-volume spectrum consists of the energy
levels of any number of noninteracting stable hadrons,
$E=\sum_{i,j}\sqrt{m_i^2+p_j^2}$, where $p_j=\tfrac{2\pi}{L}n_j$ and
$n_j$ is a vector of integers. For a nucleon at rest, the leading
excitations are states with a nucleon and any number of
pions. Positive parity requires that for an odd number of pions there
must be some nonzero momenta. These noninteracting energy gaps are
shown in Fig.~\ref{fig:dE_free}. At the physical pion mass with $m_\pi
L=4$, the lowest excitation is $\Delta E=2m_\pi$ and there are eight
$N\pi$ and $N\pi\pi$ levels below $4m_\pi$. As the box size grows, the
spectrum becomes denser.
On the other hand, at heavier pion
masses the energies rapidly increase and there are soon just a few
levels below $\Delta E=1$~GeV. Going beyond the noninteracting
approximation, in Ref.~\cite{Hansen:2016qoz} finite-volume
quantization conditions were applied in the $N\pi$ sector using the
experimentally measured scattering phase shift. Significant deviations
from the noninteracting levels were found, particularly for energy
gaps between 0.4 and 0.8~GeV, however the general features of the
$N\pi$ spectrum were not significantly changed.

Predicting excited-state contributions to two-point and three-point
functions requires knowing the spectrum $E_n$, the overlap factors
$Z_n$, and the operator matrix elements $\langle
n'|\mathcal{O}|n\rangle$. The key insight that allows for study using
chiral perturbation theory (ChPT) is that, if one assumes the smearing
size of the interpolator is small compared with $m_\pi^{-1}$, then at
leading order a single low-energy constant controls the coupling of a
local interpolator to both nucleon and nucleon-pion states, and it is
eliminated when forming ratios~\cite{Bar:2015zwa}.

The prediction from ChPT for the nucleon effective mass is a
percent-level excited-state contribution for $T\gtrsim
1$~fm~\cite{Tiburzi:2009zp,Bar:2015zwa,Tiburzi:2015tta}. On the other
hand, using the ratio method, an effect at the 10\% level is predicted
for $g_A$ at $T=1$~fm~\cite{Tiburzi:2015tta,Bar:2016uoj}. This effect
increases the effective lattice value of $g_A$, in contrast with most
numerical studies, which find that excited states decrease the
extracted value of $g_A$. At this conference, O.~Bär presented the
first study of observables at nonzero momentum transfer, namely the
form factors $G_A$ and $G_P$ of the axial current; a new tree-level
diagram was found to produce a very large excited-state effect in
$G_P$~\cite{Bar:2018akl}.

\begin{figure}
  \centering
  \includegraphics[width=0.6\textwidth]{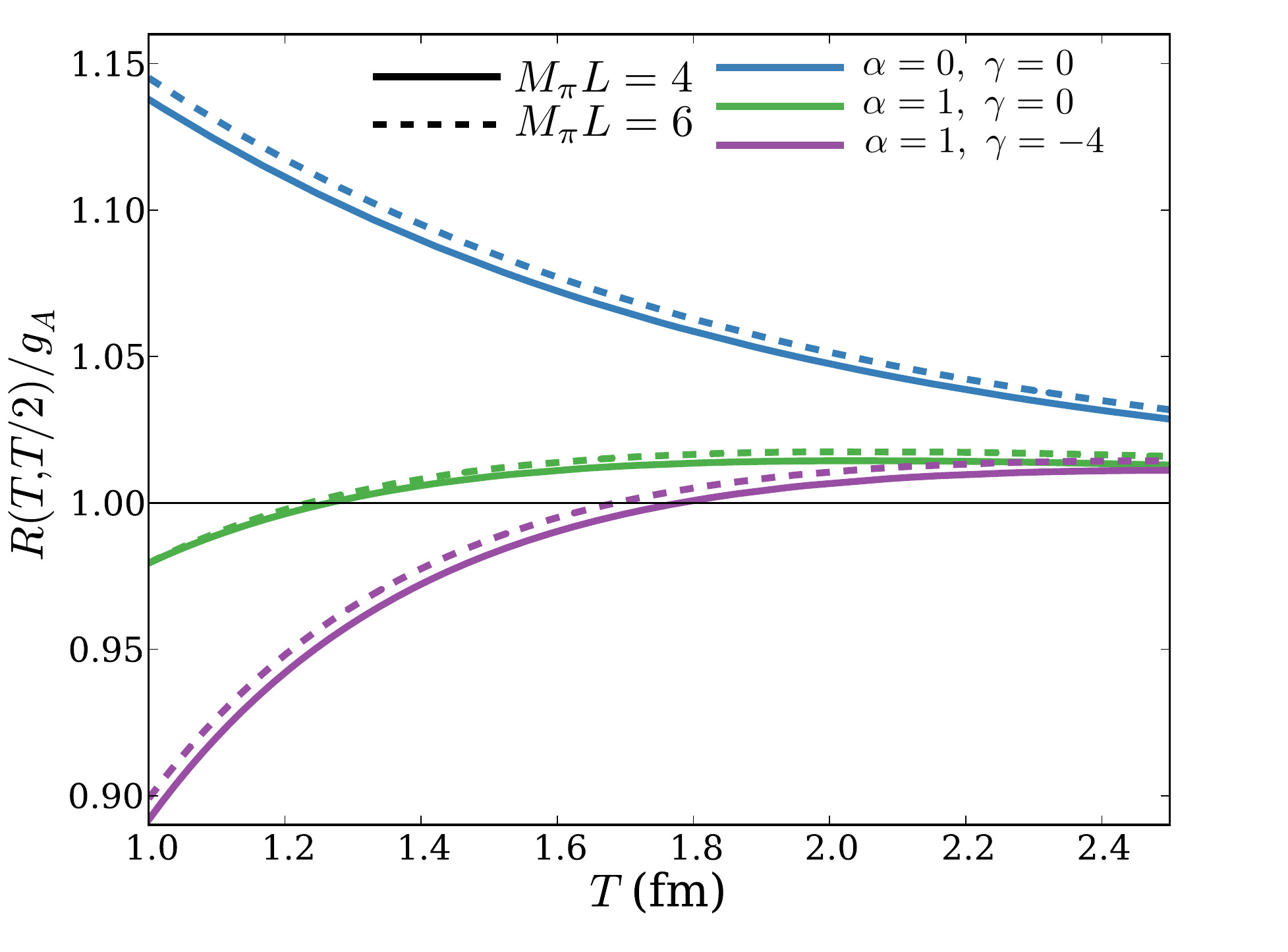}
  \caption{Model prediction of excited-state contributions to $g_A$
    determined using the ratio method~\cite{Hansen:2016qoz}. The upper
    pair of curves show the ChPT prediction, with the normalization of
    states modified using the experimental $N\pi$ phase shift and
    finite-volume quantization conditions. The other two pairs of curves show
    different scenarios for modifications to the ChPT result in the
    resonance regime. The figure is reproduced from
    Ref.~\cite{Hansen:2016qoz} under the
    {Creative Commons Attribution License}.
  }
  \label{fig:gA_HM}
\end{figure}

ChPT is not expected to produce accurate results for the contributions
from excited states with higher energies, particularly in the vicinity
of resonances; this means that it is only expected to be valid at
large source-sink separations where these contributions are
suppressed, namely $T\gtrsim 2$~fm for $g_A$ and similar
observables~\cite{Bar:2017kxh}. Deviations from ChPT for $N\pi$ states
in the resonance regime were modeled in Ref.~\cite{Hansen:2016qoz};
see Fig.~\ref{fig:gA_HM}. Under some model scenarios, the ratio-method
result for $g_A$ could be suppressed by excited states at short
source-sink separations and then rise to sit 1--2\% above the true
value for a wide range of larger source-sink separations. Clearly,
this corresponds to multiple excited states contributing with
different signs. This sort of scenario is particularly troublesome, as
it makes systematically improving a calculation by removing
excited-state effects quite difficult.

\subsection{Numerical studies}

Before discussing studies in the literature, it should be noted that
the available set of source-sink separations $T$ and source-operator
separations $\tau$ depends on how the three-point function is
computed.
\begin{compactitem}
\item The most common approach uses a \textbf{fixed
    sink}. $C_\text{3pt}(\tau,T)$ is evaluated using a sequential
  propagator through the sink, so that $T$, the sink momentum, and the
  interpolating operators are fixed. All values of $\tau$ can be
  obtained and any quark bilinear operator $\mathcal{O}(\tau)$ can be
  used. The computational cost increases with every value of $T$.
\item If one instead uses a \textbf{fixed operator}, the sequential
  propagator is evaluated through $\mathcal{O}(\tau)$, which is thus
  fixed. All values of $T$ can be obtained, and the sink momentum and
  interpolator can be varied. The computational cost increases with
  each operator insertion and with each value of $\tau$. This approach
  has been used recently in some variational studies by the CSSM
  group~\cite{Owen:2012ts,Stokes:2018emx}.
\item Rather than fixing $\tau$, one can sum over it to obtain a
  \textbf{summed operator} in the sequential propagator. In this case,
  $T$ becomes the only relevant time separation and all values can be
  obtained. The sink interpolator can be varied and the computational
  cost increases with each operator insertion. This approach has been
  used recently by CalLat~\cite{Bouchard:2016heu} and
  NPLQCD~\cite{Savage:2016kon}.
\end{compactitem}
It is possible to replace the sequential propagator with a stochastic
one. This allows for increased flexibility but at the possible cost of
increased noise~\cite{Alexandrou:2013xon, Bali:2013gxx, Yang:2015zja,
  Bali:2017mft, Gambhir_Lat18}.

\subsubsection{Improving the nucleon interpolator}

The usual\footnote{A common alternative is to replace $C\gamma_5$ with
  $C\gamma_5 P_+$, where $P_+\equiv\tfrac{1}{2}(1+\gamma_0)$ is a
  positive parity projector.} proton interpolator is
$\chi=\epsilon_{abc}(u^T_a C\gamma_5 d_b)u_c$, constructed using
smeared quark fields. Standard practice is to tune the smearing width
such that the nucleon effective mass reaches a plateau as early as
possible. It is well known from spectroscopy that the variational
method~\cite{Michael:1985ne,Luscher:1990ck}, where one finds a linear
combination of interpolating operators $\chi_\text{var}\equiv\sum_i
c_i \chi_i$ with optimized coefficients $c_i$, is a powerful
systematic approach for eliminating contributions from the
lowest-lying excited states to estimates of energy levels and matrix
elements~\cite{Blossier:2009kd,Bulava:2011yz}. In practice, its
effectiveness depends on the choice of interpolator basis
$\{\chi_i\}$.

\begin{figure}
  \centering
  \includegraphics[width=0.525\textwidth]{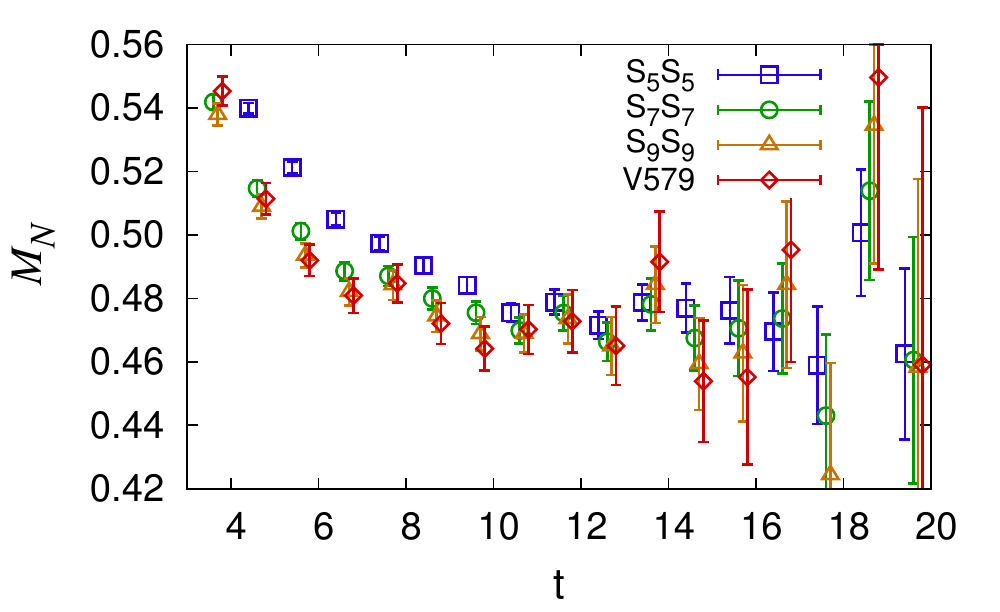}
  \includegraphics[width=0.465\textwidth]{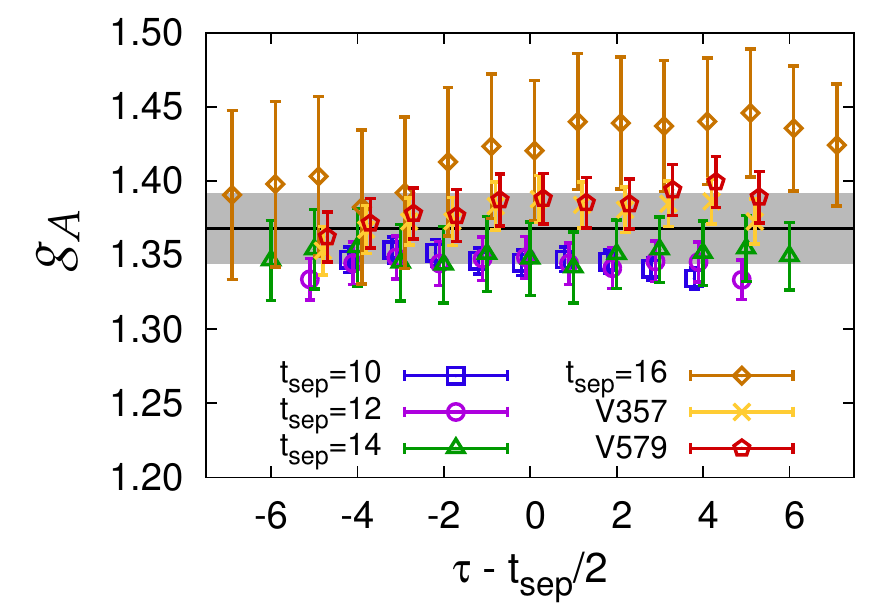}
  \caption{Comparison of analyses using a smeared standard
    interpolator (denoted S$_i$ for smearing width $i$ in lattice
    units) and using a variationally optimized interpolator (denoted
    V$ijk$, using the basis
    $\{\text{S}_i,\text{S}_j,\text{S}_k\}$). The calculation was
    performed using an ensemble with $m_\pi=312$~MeV and
    $a=0.081$~fm. Left: nucleon effective mass. Right: estimators for
    bare $g_A$, using the interpolators S$_9$ (with four source-sink
    separations), V357, and V579 (with source-sink separation
    $12a$). The plots are reproduced from Ref.~\cite{Yoon:2016dij}.}
  \label{fig:NME_variational}
\end{figure}

A simple way to produce several interpolators is to vary the smearing
width; there are two recent studies\footnote{See
  Refs.~\cite{Engel:2009nh,Owen:2012ts} for earlier studies.} that
used bases comprising interpolators with three different smearing
widths~\cite{Yoon:2016dij,Dragos:2016rtx}. Some results from
Ref.~\cite{Yoon:2016dij} are shown in
Fig.~\ref{fig:NME_variational}. The effective mass from the
variationally optimized interpolator lies very close to that of the
standard interpolator with the widest smearing. The optimized operator
and the widest smearing also both show little sign of excited-state
effects in the plateaus for the axial charge, but narrower smearings
do show clear signs (see e.g.\ Fig.~15 of
\cite{Yoon:2016dij}). Ref.~\cite{Dragos:2016rtx} also found that
narrower smearings suffer from larger excited-state effects. In that
study, the variationally optimized interpolator produced smaller
excited-state effects than the largest smearing. However, a still
larger smearing might be as good as the variational interpolator. One
should take two clear lessons from these studies. The first is that
tuning the smearing width is important, since it can have a
significant effect on excited states. The second is that when studying
a computationally more expensive alternative to the standard approach,
a fair comparison should be with a well-tuned operator; it is easy to
make the standard approach appear to be worse by using a too-narrow smearing.

Beyond varying the smearing width, one can use different local
operator structures such as $\epsilon_{abc}(u_a^TCd_b)\gamma_5
u_c$. This approach has been used to add negative-parity interpolators
to the basis, which can be important for coupling to excited states in
moving frames~\cite{Stokes:2018emx}. Including covariant derivatives
or the chromomagnetic field strength in the interpolator allows for a
much larger basis~\cite{Edwards:2011jj,Dudek:2012ag}. In
Ref.~\cite{Egerer:2018xgu}\footnote{C.~Egerer presented preliminary
  results at this conference.}, such a basis was used, employing the
distillation method to efficiently construct the correlators. In
general, a larger basis was more effective at removing excited-state
contributions, with the largest effect seen in the tensor charge. This
reference also reports that the Laplacian-Heaviside smearing used in
distillation produces smaller excited-state effects from the standard
operator than the more commonly used Wuppertal smearing, however it is
unclear whether either smearing was tuned.

\subsubsection{Fitting excited states}

A natural strategy for removing contributions from excited states is
to fit correlators [or derived quantities such as $R(\tau,T)$ or
$S(T)$] using a model that includes excited-state effects. By far the
most common model is based on a truncation of the spectral
decomposition to a small number of states (two or three). Generally
the two-point function provides the strongest constraints on the
energies $E_n$ and overlaps $Z_n$, and the three-point function serves
to determine the matrix elements $\langle n'|\mathcal{O}|n\rangle$.
Commonly, the energy gap $\Delta E$ is found to be between 0.5 and
1.0~GeV, which is usually greater than the expected lowest-lying
excitation energy shown in Fig.~\ref{fig:dE_free}.

\begin{figure}
  \centering
  \includegraphics[width=0.495\textwidth]{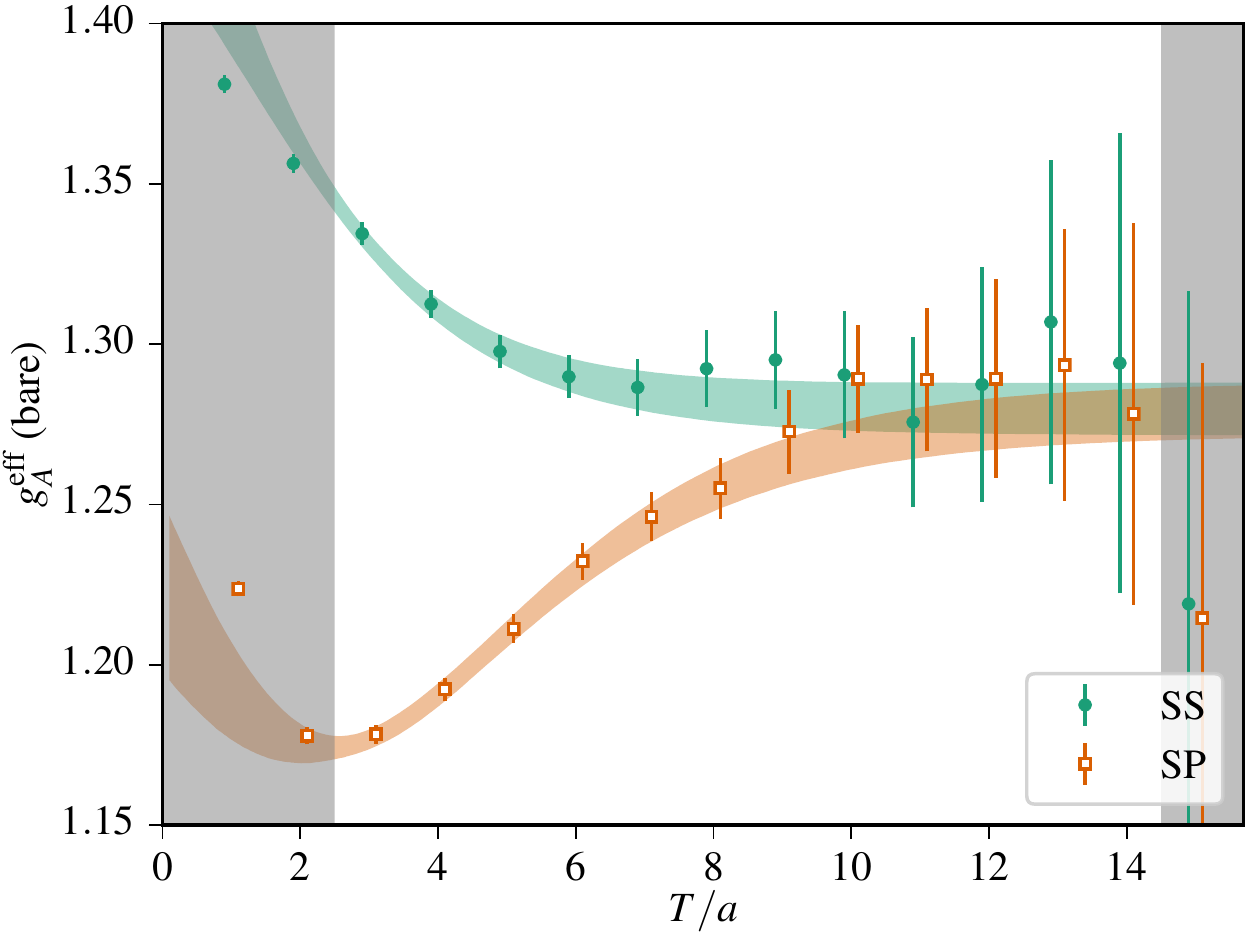}
\begin{minipage}[b]{0.495\textwidth}
\begin{center}
  \includegraphics[width=\textwidth]{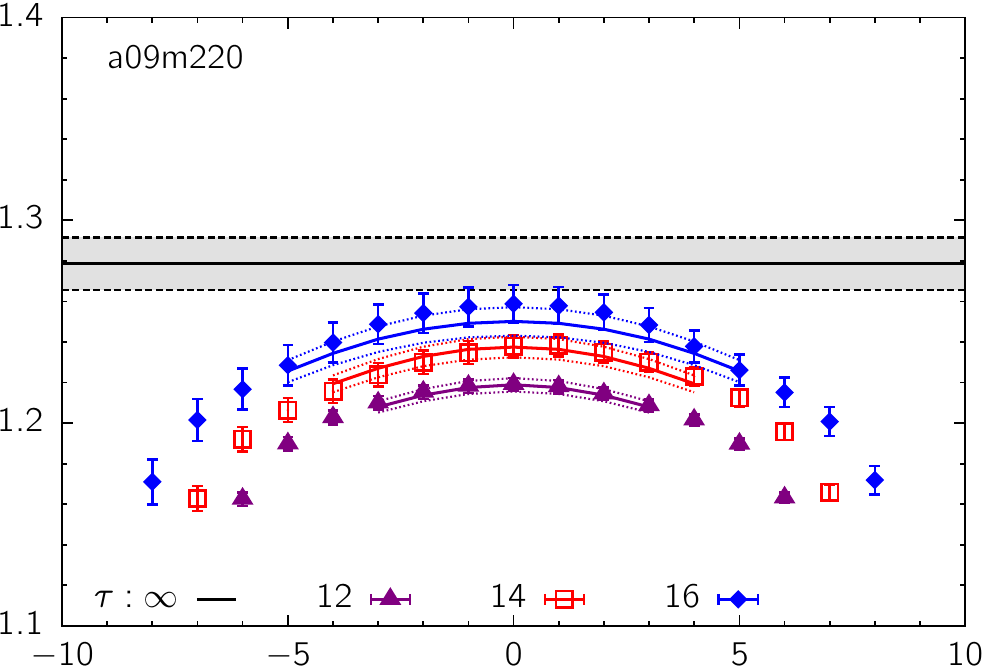}\\[-0.9ex]
  {\scriptsize $(\tau-T/2)/a$}
\end{center}
\end{minipage}
  \caption{Excited-state fits for bare $g_A$. Both calculations were
    performed using the same HISQ ensemble with $a\approx 0.09$~fm and
    $m_\pi\approx 220$~MeV. Left: two-state fit to data using domain
    wall valence quarks and the summed operator method, with smeared
    and point sinks (filled circles and open squares,
    respectively). Right: three-state fit to data with clover valence
    quarks and the fixed sink method, for three source-sink
    separations. The left plot is adapted from
    Ref.~\cite{Chang:2018uxx} and the right plot is reproduced from
    Ref.~\cite{Gupta:2018qil} under the Creative Commons Attribution
    License.}
  \label{fig:gA_fits}
\end{figure}

Some recent examples are given in Refs.~\cite{Chang:2018uxx,
  Gupta:2018qil}. In Ref.~\cite{Chang:2018uxx}, the extracted value of
$g_A$ was shown to be stable when the fit range is varied; similarly,
in Ref.~\cite{Gupta:2018qil} stability was shown with respect to
varying fit ranges and the number of states in the fit
model. Figure~\ref{fig:gA_fits} shows the preferred fits for $g_A$ on
the same ensemble (albeit with different valence quark action). In
both cases, the fits start with a minimum time separation of
$3a$. Given the pion mass and box size, there are more than ten
noninteracting excited energy levels with energy gap $\Delta E<1$~GeV;
for the first points included in the fit these are only suppressed by
$e^{-3a\Delta E}>0.25$. Clearly, it is difficult to associate the
``excited state'' in the fit with a single actual state.

At this conference, K.~Ottnad presented a different fitting
approach~\cite{Ottnad:2018fri} that does not determine the energy gap
$\Delta E$ from the two-point function. Instead, fits are performed to
the ratios $R(\tau,T)$; fitting six different observables
simultaneously is sufficient to constrain $\Delta E$. Interestingly,
in this case the fitted energy gap approaches the expected
lowest-lying noninteracting level as the minimum time separation
included in the fit, $t_\text{start}$, is increased.

In order for these fits to be trustworthy, ideally they would be
required to have good fit qualities ($p$-values). (This is not a
sufficient condition!)  A strong test is given in
Ref.~\cite{Borsanyi:2014jba}: when the fit is repeated on many
ensembles, the distribution of fit qualities should be uniform, which
can be checked using a Kolmogorov-Smirnov (KS) test. More than half of
the fit qualities in Ref.~\cite{Chang:2018uxx} are below 0.2, and
hence the KS test indicates that they are not compatible with the
uniform distribution ($p<10^{-3}$). In contrast, the fits in
Ref.~\cite{Gupta:2018qil} are acceptable from this point of view
($p=0.26$). Some caution may be required, however, when fitting to
many variables, as the difficulty in inverting a large covariance
matrix can make it difficult to reliably estimate $\chi^2$ and the fit
quality.

\subsection{Outlook on excited states}

It is natural to ask why the picture in Fig.~\ref{fig:dE_free} of
low-lying $N\pi$ and $N\pi\pi$ states does not appear in the spectrum
of typical variational analyses or multi-state fits. One argument is
that the coupling of a multiparticle state to a local interpolator is
suppressed by the inverse lattice volume. However, this should be
compensated by the density of states so that in infinite volume the
interpolator will couple to continua of multiparticle states. In fact,
model predictions such as Fig.~\ref{fig:gA_HM} show a weak volume
dependence; this suggests that continuum spectral functions might
yield suitable fit models for excited states in large volumes.

The absence of multiparticle states is familiar from meson
spectroscopy. It has been found that a variational basis must include
nonlocal operators in order to identify the complete and correct
spectrum; see, e.g., \cite{Wilson:2015dqa}. Even though nucleon
structure only requires removing excited states and not obtaining
precise knowledge of them, it may still be necessary to include
nonlocal operators in order to benefit in practice from the proven
improved asymptotic approach to the ground state when using the
variational method~\cite{Blossier:2009kd,Bulava:2011yz}.

Given that state of the art nucleon structure calculations have not
reconciled their data with theoretical expectations for excited-state
effects, perhaps the safest approach is to analyze data in multiple
ways: ratio and summation methods, which don't make specific
assumptions about the spectrum of excitations, as well as fits that
can make use of shorter time separations. This was done in the
extensive excited-state study of Ref.~\cite{Dragos:2016rtx}, which
also included a variational setup, as well as in some recent
physical-pion-mass calculations by ETMC (see e.g.\
\cite{Alexandrou:2017hac}).

\section{Finite-volume effects}\label{sec:volume}

There is a long history of attributing a low value for $g_A$ computed
on the lattice to finite-volume effects. These effects were computed
in ChPT in Ref.~\cite{Beane:2004rf}; neglecting loops with $\Delta$
baryons, the leading contribution at large volume is
\begin{equation}\label{eq:gA_volume}
  \frac{g_A(L)-g_A}{g_A} \sim \frac{m_\pi^2 g_A^2}{\pi^2F_\pi^2}
  \sqrt{\frac{\pi}{2m_\pi L}} e^{-m_\pi L}.
\end{equation}
In addition to being exponentially suppressed at large $m_\pi L$, if
one fixes $m_\pi L$ and decreases $m_\pi$ this effect will also be
reduced. If this expression holds true, then a calculation with $m_\pi
L=3$ at the physical pion mass will have smaller finite-volume effects
than one with $m_\pi L=4$ at $m_\pi=300$~MeV.

\begin{figure}
  \centering
  \includegraphics[width=\textwidth]{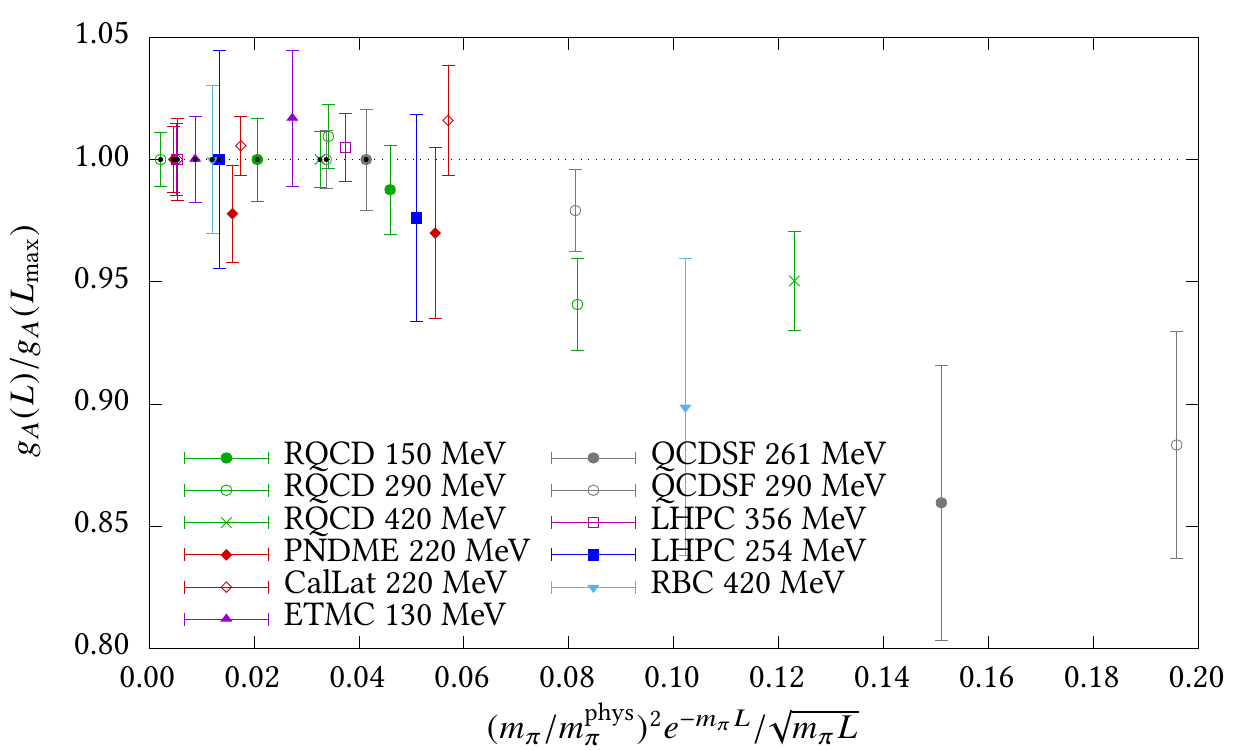}
  \caption{Controlled studies of finite-volume effects in
    $g_A$~\cite{Bali:2014nma, Gupta:2018qil, Chang:2018uxx,
      Lauer_Lat18, Horsley:2013ayv, Bratt:2010jn, Green:2013hja,
      Yamazaki:2008py}. Each study has two or more volumes at the same
    pion mass, which is indicated in the legend. The largest volume
    of each study is used for normalization, and a black dot is placed
    over the symbol to indicate that the central value is fixed at
    one. The horizontal axis contains the dependence on pion mass and
    volume from Eq.~\protect\eqref{eq:gA_volume}.}
  \label{fig:gA_volume}
\end{figure}

There have been several fully-controlled studies of finite-volume
effects in $g_A$, i.e., the same calculation performed on ensembles
that differ only by their volume. These are
summarized\footnote{Calculations where the lattice temporal extent was
  varied together with the spatial volume are also included.} in
Fig.~\ref{fig:gA_volume}. For large values of $m_\pi L$ and small pion
masses, no effect is observed within uncertainties at the few percent
level. As the right hand side of Eq.~\eqref{eq:gA_volume} is
increased, the first significant effect is at the 5\% level in the
calculation by RQCD at $m_\pi=290$~MeV and $m_\pi L =3.4$. However, this
is a negative effect rather than the positive effect predicted by
Eq.~\eqref{eq:gA_volume}. Encouragingly, the physical pion mass with
$m_\pi L=3$ corresponds to 0.03 on the horizontal axis, where no
effect has been detected.

Global fits to a set of ensembles --- where the pion mass, lattice
spacing, and volume are all varied --- provide a different approach to
study finite-volume effects. The challenge is that any failure of the
fit function to accurately describe the dependence on the other
variables is a source of systematic uncertainty in the estimate of
finite-volume effects. This is especially true because most sets of
ensembles will tend to have larger values of $m_\pi L$ at larger pion
masses and on coarser lattice spacings. Using a global fit,
Ref.~\cite{Gupta:2018qil} assumed the volume dependence of $g_A$ has
the form $cm_\pi^2 e^{-m_\pi L}$ with $c$ a free parameter, and found
a $-0.9(5)\%$ effect at the physical pion mass with $m_\pi L=4$. A
similar approach was used by Ref.~\cite{Ottnad:2018fri}, and a similar
effect size was found. Finally, Ref.~\cite{Chang:2018uxx} assumed the
leading heavy baryon ChPT expression and allowed a higher-order term
proportional to $m_\pi^3$; this also produced a small effect at the
physical pion mass.

\section{Chiral extrapolation}\label{sec:chiral}

In heavy baryon ChPT, the pion mass dependence of the axial charge is
known to take the form
\begin{equation}
  g_A(m_\pi) = g_0 - (g_0 + 2g_0^3)\left(\frac{m_\pi}{4\pi F_\pi}\right)^2\log\frac{m_\pi^2}{\mu^2} + c_1 m_\pi^2 + c_2 m_\pi^3 + O(m_\pi^4),
\end{equation}
where $g_0$ is the axial charge in the chiral limit and $c_{1,2}$ are
additional low-energy constants. Although the prefactor of the chiral
log is known, it is unclear how high of a pion mass can be reached
before the convergence of ChPT breaks down. As a result, recent
calculations that performed a chiral fit~\cite{Capitani:2017qpc,
  Yamanaka:2018uud, Chang:2018uxx, Liang:2018pis, Gupta:2018qil,
  Ottnad:2018fri}, including results presented by R.~Gupta and
K.~Ottnad at this conference, generally preferred to use a simple
polynomial dependence on $m_\pi^2$.

The issue of chiral extrapolation can, of course, be avoided by using
only lattice ensembles with near-physical pion masses. Results using
this approach were presented at this conference by
M.~Constantinou~\cite{Constantinou_Lat18},
Y.~Kuramashi~\cite{Shintani:2018ozy}, C.~Lauer~\cite{Lauer_Lat18},
Y.~Lin~\cite{Lin_Lat18}, and S.~Ohta~\cite{Ohta:2018zfp}.

\begin{SCfigure}
  \includegraphics[width=0.7\textwidth]{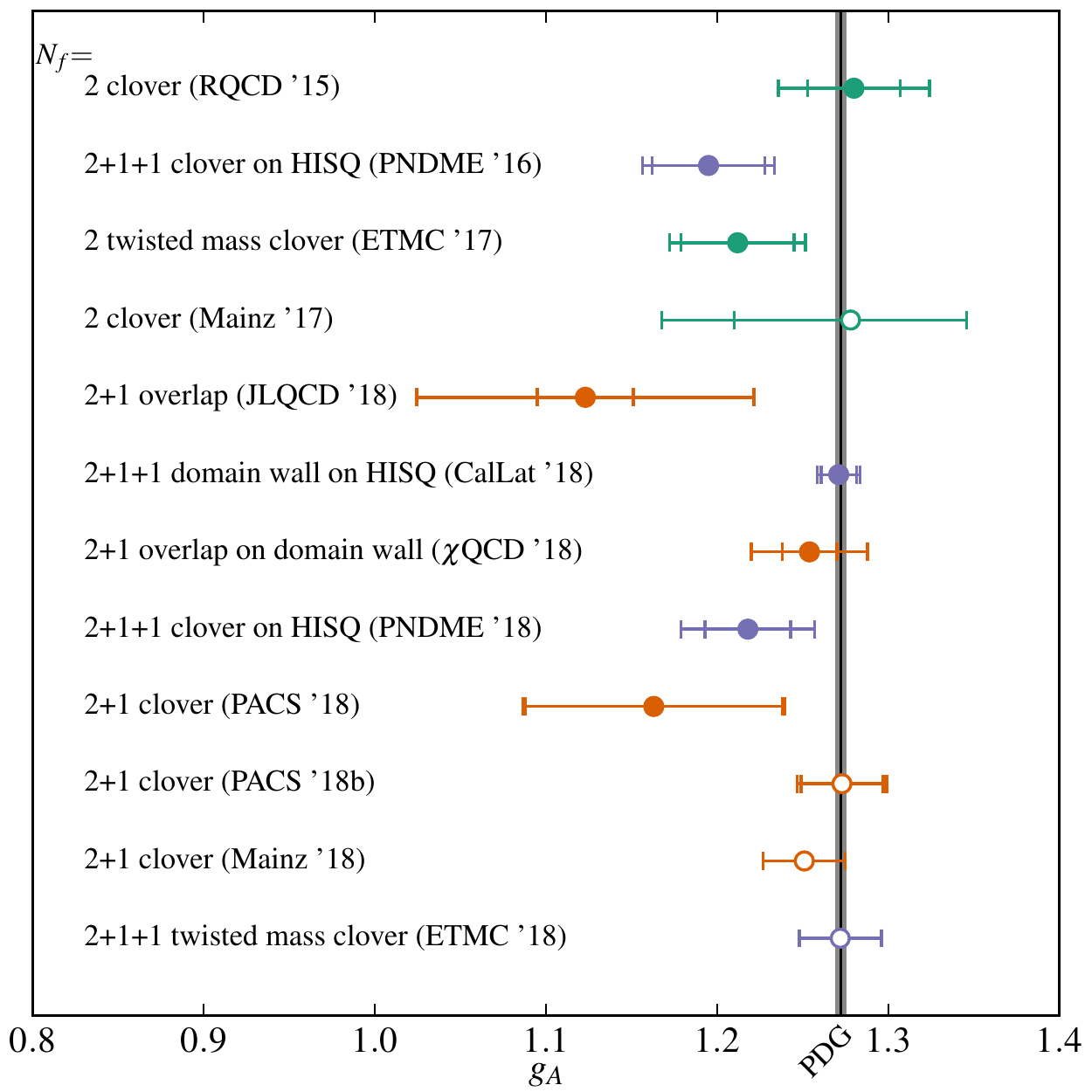}
  \caption{Recent calculations of the nucleon axial charge using $2$
    flavours of dynamical sea quarks~\cite{Bali:2014nma,
      Alexandrou:2017hac, Capitani:2017qpc} (green), $2+1$
    flavours~\cite{Yamanaka:2018uud, Liang:2018pis, Ishikawa:2018rew,
      Shintani:2018ozy, Ottnad:2018fri} (orange), and $2+1+1$
    flavours~\cite{Bhattacharya:2016zcn, Chang:2018uxx, Gupta:2018qil,
      Constantinou_Lat18} (blue). Published results are shown with a
    filled symbol. The PDG value~\cite{Tanabashi:2018oca} is indicated
    by the vertical band.}
  \label{fig:gA_results}
\end{SCfigure}

Figure~\ref{fig:gA_results} shows some recent determinations of the
axial charge. It is encouraging that several collaborations are now
able to reproduce the experimental value, although there is still a
tendency for results to sit below experiment and no result is even
half a standard deviation above experiment. Given that the
experimental value is known, it may be useful for future calculations
to perform a blinded analysis.

\section{Summary and outlook}\label{sec:outlook}

Excited-state contamination remains a major focus of nucleon structure
calculations. A full variational study including nonlocal
interpolators that couple well to multiparticle states could help to
determine whether current methods are adequate. In contrast, no sign
of large finite-volume effect in $g_A$ has been observed in the
existing fully-controlled studies at low pion masses.

Discretization effects were not discussed in this review, in part
because they are not universal. In general they appear to be less
important than excited states, but they are nevertheless important for
controlling uncertainties and can have a significant impact on the
outcome such as in Ref.~\cite{Gupta:2018qil}.

It should be stressed that the results in Fig.~\ref{fig:gA_results}
have not been filtered based on any quality criteria. Such an
evaluation is necessary for obtaining a reliable ``lattice QCD
average'' of any observable. However, it may be particularly difficult
to set standards for controlling excited-state effects, since analysis
strategies vary significantly. A first community attempt was made in
Ref.~\cite{Lin:2017snn}, and some nucleon structure will also be
included in the next FLAG review.

Bringing simple observables like $g_A$ under precise control over all
sources of systematic uncertainty will be an important step toward
reliable calculations of more complex observables such as the proton
charge radius and parton distribution functions. Systematics in these
observables will require further study; in particular, finite-volume
effects could be important for form factors~\cite{Shintani:2018ozy}
and discretization effects might be significant for parton
distribution functions, which are generally not $O(a)$ improved.

\acknowledgments

I thank everyone who sent results in advance of my talk and who
replied to my questions: C.~Alexandrou, C.~C.~Chang, C.~Egerer,
R.~Gupta, J.~Liang, S.~Ohta, K.~Ottnad, F.~M.~Stokes, A.~Walker-Loud,
and T.~Yamazaki. I also thank my colleagues at DESY for their comments
on an early version of this talk and Karl Jansen for sending comments
on a draft of these proceedings.

\bibliography{proceedings}
\bibliographystyle{utphys-noitalics}

\end{document}